\documentclass[final,authoryear,3p,times]{elsarticle}
\usepackage{amsmath,amsfonts,amssymb}
\usepackage{graphicx,amsmath}  
\usepackage{epstopdf}    
\usepackage{relsize}      
\usepackage{color}
\usepackage{MnSymbol}
\usepackage{subfigure}
\def\micron{\:\mu\mbox{m}}

\newcommand{\rw}{\rightarrow}
\newcommand{\bea}{\begin{eqnarray}}
\newcommand{\eea}{\end{eqnarray}}

\newcommand{\cf}{{\it cf.}~}
\newcommand{\etc}{{\it etc.}~}

\newcommand{\nm}{\:\mbox{nm}}
\def\half{\frac{1}{2}}

\journal{Journal of the Mechanics and Physics of Solids}

\begin{document}

\begin{frontmatter}

\title{Obstacles and sources in dislocation dynamics: Strengthening and statistics of abrupt plastic events in nanopillar compression}
\author[JHU]{S. Papanikolaou \corref{cor} }
\author[RuG]{H. Song}
\author[RuG]{E. Van der Giessen \corref{cor} }
\cortext[cor]{Corresponding authors}
\address[JHU]{Department of Mechanical Engineering, The Johns Hopkins University, 3400 N Charles St, Baltimore, Maryland, 21218}
\address[RuG]{Zernike Institute for Advanced Materials, University of Groningen, 9747 AG Groningen, the Netherlands}
\ead{spapanikolaou@jhu.edu, E.van.der.Giessen@rug.nl}

\begin{abstract}
Mechanical deformation of nanopillars displays features that are distinctly different from the bulk behavior of single crystals: Yield strength increases with decreasing size and plastic deformation comes together with strain bursts or/and stress drops (depending on loading conditions) with a very strong sensitivity of the stochasticity character on material preparation and conditions. The character of the phenomenon is standing as a paradox: While these bursts resemble the universal, widely independent of material conditions, noise heard in bulk crystals using acoustic emission (AE) techniques, they strongly emerge primarily with decreasing size and increasing strength in nanopillars. 
In this paper, we present a realistic but minimal discrete dislocation plasticity model for the elasto-plastic deformation of nanopillars that is consistent with the main experimental observations of nano pillar compression experiments and provides a clear insight to this paradox.  With increasing sample size, the model naturally transitions between the typical progressive behavior of nanopillars to a behavior that resembles evidence for bulk mesoscale plasticity. The combination of consistent strengthening, large flow stress fluctuations and critical avalanches is only observed in the  {\it depinning regime} where obstacles are much stronger than dislocation sources; in contrast, when dislocation source strength becomes comparable to obstacle barriers, then yield strength size effects are absent but plasticity avalanche dynamics is strongly universal, across sample width and aspect-ratio scales. Finally, we elucidate the mechanism that leads to the connection between depinning and size effects in our model dislocation dynamics. In this way, our model builds a way towards unifying statistical aspects of mechanical deformation across scales.

\end{abstract}

\end{frontmatter}

\section{Introduction}
The dynamical character of crystal plasticity at the nanoscale has been under scrutiny for more than a decade\newline ~\citep{Uchic03,Uchic04,Uchic05,Dimiduk2006,Uchic2009, Kraft2010,Greer2011}. This interest is driven by the identification of unconventional size effects in samples made by the focused ion beam technique under uniaxial tension or compression. Experiments of nanocrystalline pillar tension and compression have convincingly shown apparent strengthening with decreasing pillar width $w$, with the yield strength varying as $\sigma_Y\sim w^{-n}$ with $n\in(0.4,0.8)$~\citep{Uchic2009,Greer2011}, and a mild decrease with slenderness $\alpha=h/w$~\citep{volkert2006,kiener2008,Senger2011}. The mechanism of strengthening in nanopillars has been attributed to a transition from typical Frank-Read sources in the bulk to the predominance of atypical sources such as surface sources~\citep{gall2004,park2006,diao2006} and single-arm sources~\citep{Weinberger2008,oh2009,zheng2010,lu2010, Ryu2015} at the nanoscale~\citep{Jennings2011,Greer2011,Uchic2009,BulatovCaiBook, Parthasarathy2007}. 

Nano-strengthening is accompanied by large, abrupt strain jumps (load control) and/or stress drops (displacement control)~\citep{Dimiduk2006,Uchic04,Dimiduk2005,Dimiduk2007,Uchic05,Greer2005,Greer2006,Ng2007,Ng2008,Ng2008b,18}. The stochastic abrupt events resemble noise/avalanches in disordered magnets or earthquakes~\citep{2,Dimiduk2006,Uchic05,Papanikolaou2012, P18}. Analysis of the statistics of abrupt plastic events has revealed that nanopillar events, statistically, appear to follow power-law-tailed distributions for strain steps with a large event cutoff that depends on specimen width~\citep{2,Miguel2001b, Weiss2003,miguel2001}. The actual nature of these events has remained somewhat elusive, partially due to the complexity of loading paths, intertwining slip and stress events. However, AE measurements in a multitude of materials~\citep{weiss15, Miguel2001b} have shown the presence of ubiquitous power-law plastic events that are independent of loading paths as well as sample dimensions. The energy release during such bulk events statistically displays a  power law distribution with exponent $\tau\in (1.4, 1.9)$ and no apparent cutoff dependence on sample parameters~\citep{weiss15}; in nanopillars, the analogous exponent range is $\tau\in(1.3, 2.1)$ for strain jumps or stress drops (depending on selected loading path), and it shows strong fluctuations with sample dimensions~\citep{Greer2011}.

The apparent contrast between bulk and nanoscale mechanical behavior becomes almost paradoxical when it is considered that the ubiquitous nano-strengthening has not been observed in bulk strength measurements; thus, this might suggest that there is no relation between nano-strengthening and plastic avalanche statistics. On this basis, theoretical studies of the stochastic/abrupt plastic flow have mainly adopted continuum methods~\citep{alava2014,Miguel2001b,zaiser,koslowski,Papanikolaou2012} thus addressing bulk properties, or two-dimensional discrete dislocation dynamics in the special environment of randomly placed edge dislocations in periodic systems with no obstacles or dislocation sources~\citep{miguel2001,Ispanovity2013,Ispanovity2010, Zaiser2007,alava2014}. Such approaches have provided useful insights for predicting unconventional power-law statistics of abrupt plastic events in crystals~\citep{Weiss2003}, but they lack typically considered mechanisms that give rise to nano-strengthening (starvation, source/obstacle exhaustion, single-arm source proliferation \etc). 

Dislocation sources can be either bulk --typically of the Frank-Read type~\citep{1}-- or unconventional, such as surface and single-arm sources~\citep{Uchic2009}; it is believed that bulk sources typically require much smaller (by an order of magnitude) stress to be activated than unconventional ones~\citep{Greer2011}. Dislocation obstacles capture the effect caused by precipitates, as well as forest dislocations that cross the glide slip planes. As molecular (MD)  and three-dimensional discrete dislocation (3D-DDD)  simulations have been showing~\citep{Madec2002}, the strength of such obstacles varies strongly with dislocation configurations. Actual statistical properties of dislocation obstacle distributions currently remain unknown and thus, it is currently not feasible to investigate, using MD or 3D-DDD, the effect of obstacles on dislocation dynamics. However, two-dimensional discrete edge dislocation (2D-DDD) simulations become ideal for the investigation of various obstacle-related statistical properties, since obstacles are minimally and randomly introduced with a pre-defined statistical distribution. Indeed, \cite{curtin} have developed a direct connection between the yield strength of a system of edge dislocations in single slip and obstacle spacing, obstacle strength, source nucleation strength and average source spacing. Thus, it appears critical to investigate the role of the ratio of source to obstacle strengths for nanopillar strengthening and plastic flow avalanche dynamics.

In this paper, we propose a realistic but minimal 2D discrete dislocation plasticity model for elastoplastic deformation of nanopillars and show that its predictions are consistent with the main experimental observations of nanopillar compression experiments in terms of strengthening as well as stochastic plastic flow. The explanation of the apparent paradox mentioned above naturally emerges through the stochastic competition among dislocation sources and obstacles.

The remainder of this paper is organized as follows:
Section 2 describes the methodology of our 2D-DDD model that is built to phenomenologically but minimally capture the details of nanopillar compression experiments, as well as the details of our statistical analysis of dislocation dynamics, similarly to common experimental protocols. Our original model contains large-activation-stress surface and small-activation-stress bulk sources, as well as quite strong obstacles -- all at fixed sample densities.
Section 3 is focused on experimentally relevant results of our 2D-DDD model simulations, such as the behavior of the yield stress as function of sample width and aspect ratio, or in terms of other material parameters. In addition, we present the statistical aspect of pillar compression beyond yielding and we show qualitative agreement with  uniaxial nanopillar compression experiments. 
Furthermore, we investigate the role of surface sources for our strengthening and statistics results, by performing extensive simulations in a simplified model where surface dislocation sources are absent, but bulk properties remain identical with our original 2D-DDD model. 
In Section 4, we investigate in detail the relative role of source strength to obstacle strength in our simplified model, showing the two qualitatively different (in strengthening and statistics) regimes that we fully characterize. In Section 5, we summarize  our results.
In Appendix A, we briefly discuss a bending effect that emerges during pillar compression.
In Appendix B, we show that the inclusion of 2 slip systems does not qualitatively alter our main conclusions.

\section{Model description and Methods}
\label{sec:model}
\begin{figure}[ht!]
\centering
\includegraphics[width=0.7\textwidth]{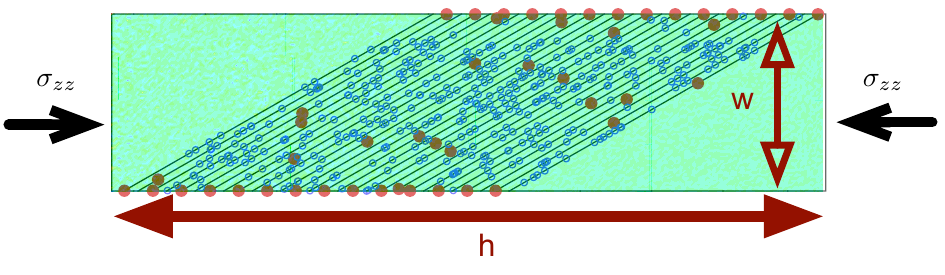}
\caption{The 2D discrete dislocation plasticity model of pillar compression: Slip planes (lines) span the sample, equally spaced at $d=10b$, where those close to corner slip plane are considered inactive to maintain a smooth loading boundary. Surface and bulk dislocation sources are present (disks) and forest obstacles are spread homogeneously across the active slip planes. Initially the sample is stress and dislocation free. } 
 \label{fig:schematic}
\end{figure}

The geometry of the model problem is shown in Fig.~\ref{fig:schematic}. Pillars are modeled by a rectangular profile of width $w$ and aspect ratio $\alpha$ ($\alpha=h/w$). 
Plastic flow occurs by the nucleation and glide of edge dislocations, for simplicity, on a single slip system. We confine attention to edge dislocations, not only because they provide the essential aspects of crystal plasticity (Burgers vector and line direction) with no numerical overhangs, but also because most results for collective plasticity avalanche dynamics originate from or are inspired by studies of edge dislocations~\citep{Miguel2001b,Zaiser2007,Ispanovity2010,Ispanovity2013}. With the typical Burgers vector of FCC crystals being $b=0.25\nm$, we study sample widths ranging in powers of 2 from $w=0.0625$ to $1\micron$ with $\alpha=h/w=4$ to $32$. The lateral edges ($x=0,w$) are traction free, allowing dislocations to exit the sample. Loading is taken to be ideally strain-controlled, by prescribing the $z$-displacement at the top and bottom edges ($z=0,h$). The applied strain rate, $\dot{h}/h=10^4 \text{s}^{-1}$, is held constant across all our simulations, similar to experimental practice.
Plastic deformation of the crystalline samples is described using the discrete dislocation framework for small strains~\citep{vandergiessen1995},
where the determination of the state in the material employs superposition. 
As each dislocation is treated as a singularity in a linear elastic background solid with Young's modulus $E$ and Poisson ratio $\nu$, 
whose analytical solution is known at any position, this field needs to be corrected by a smooth image field $(\hat{\ })$ to ensure that actual boundary conditions are satisfied. 
Hence, the displacements $u_i$, strains $\varepsilon_{ij}$, and stresses $\sigma_{ij}$ are written as  
\begin{equation}
\label{eq:superposition}
u_i = \tilde{u_i}+\hat{u_i}, \; \varepsilon_{ij} = \tilde{\varepsilon}_{ij}+\hat{\varepsilon}_{ij}, \;  \sigma_{ij} = \tilde{\sigma}_{ij}+\hat{\sigma}_{ij},
\end{equation}
where the ($\tilde{\ }$) field is the sum of the fields of all $N$ dislocations in their current positions, i.e.
\begin{equation}
 \label{eq:dislocation-field}
\tilde{u}_i=\sum_{J=1}^{N}\tilde{u}_i^{(J)}, \; \tilde{\varepsilon}_{ij}=\sum_{J=1}^{N}\tilde{\varepsilon}_{ij}^{(J)}, \; \tilde{\sigma}_{ij}=\sum_{J=1}^{N}\tilde{\sigma}_{ij}^{(J)}.
\end{equation}
Image fields are obtained by solving a linear elastic boundary value problem using finite elements with the boundary conditions changing as the dislocation structure evolves under the application of mechanical load.

The slip spacing in this paper is different from existing 2D-DDD model studies where larger slip spacing (such as $100b$ or $200b$) is used~\citep{curtin,deshpande2006}. The possible consequence of larger slip spacing is that interactions among neighboring slip planes are weakened. 
In our model, available (but not necessarily active) slip planes are $10b$ apart and are oriented at $30^\circ$ away from the loading direction (Fig.~\ref{fig:schematic}).
At the beginning of the calculation, the crystal is stress and dislocation free. This corresponds to a well-annealed sample, yet with pinned dislocation segments left that can act either as dislocation sources or as obstacles. Dislocations are generated from sources when the resolved shear stress $\tau$ at the source location is sufficiently high ($\tau_{\rm nuc}$) for a sufficiently long time ($t_{\rm nuc}$). We consider bulk sources \citep{vandergiessen1995}, as well as surface sources. 

Each sample contains a random distribution of forest dislocation obstacles, surface dislocation sources, as well as a random distribution of bulk dislocation sources. The bulk sources are randomly distributed over slip planes at a density $\rho^{\rm{bulk}}_{\rm{nuc}}= 60 \micron^{-2}$, while their strength is selected randomly from a Gaussian distribution with mean value $\bar{\tau}_{\rm nuc} = 50$ MPa and 10\% standard deviation. Bulk sources are designed to mimic the Frank-Read mechanism in two dimensions~\citep{1}, such that they generate a dipole of dislocations at distance $L_{\rm nuc}$, when activated. The initial distance between the two dislocations in the dipole is
\begin{equation}
\label{eq:L_nuc}
L_{\rm nuc}= \frac{E}{4\pi(1-\nu^2)}\frac{b}{\tau_{\rm nuc}},
\end{equation}
at which the shear stress of one dislocation acting on the other is balanced by the local shear stress.
We only consider glide of dislocations. The evolution of the dislocation is determined by the component of the Peach-Koehler force in the slip direction. For the $I$th dislocation, this is given by
\begin{equation}
\label{eq:P-K}
f^{(I)} = \boldsymbol{n}^{(I)}\cdot\left(\boldsymbol{\hat{\sigma}}+\sum_{J\neq I}{\boldsymbol{\tilde{\sigma}}}^{(J)}\right)\cdot\boldsymbol{b}^{(I)},
\end{equation}
where $\boldsymbol{n}^{(I)}$ is the slip plane normal and $\boldsymbol{b}^{(I)}$ is the Burgers vector of dislocation $I$.
This force will cause the dislocation $I$ to glide, following over-damped dynamics, with velocity
\begin{equation}
\label{eq:B-v}
v^{(I)} = \frac{f^{(I)}}{B},
\end{equation}
where $B$ is the drag coefficient. In this paper, its value is taken as $B=10^{-4}$Pa s, which is representative for aluminum.

In addition to bulk dislocation sources, surface dislocation sources are successively placed at opposite ends of slip planes, which for the current slip plane spacing corresponds to a surface density of around $\rho^{\rm surf}_{\rm{nuc}}=175/\micron$. We find that the source density dependence of our results is small for the widths and aspect ratios considered. Surface nucleation typically takes place at much higher stress than bulk nucleation, but the extent of possible mechanisms for dislocation nucleation from the surface is not yet clear, compared to typical bulk F-R sources for 2D simulations. 3D-DDD simulations~\citep{Ryu2015} based the nucleation rate of surface sources on evidence from atomistic models: 
Once surface nucleation happens (only allowed to occur on the $\half<1 1 0>\{1 1 1\}$ type slip systems), the slip system with maximum $PK$ force on the dislocation segment closest to the cylinder axis of the pillar is selected for dislocation nucleation, and then a half loop with radius of 50$b$ is created and restricted to move inside the pillar. In our 2D model, we assume the probability of surface nucleation to depend on the assumed surface source strength; once a single dislocation is generated, it is put at $10b$ from the free surface, and will move according to the actual $PK$ force. In the absence of interactions from any other dislocations, and in order for surface dislocation to move into the pillar, the $PK$ force has to surpass the image stress of 312 MPa when the dislocation is put at $10b$ from the free surface, otherwise dislocation will be attracted to the free surface and escape. Under this circumstance, our surface nucleated dislocation has an effective nucleation strength of 312 MPa which exceeds the bulk source strength by approximately an order of magnitude. The applicability of this assumption will be verified in the subsequent sections by comparing our simulation results to typical experimental phenomenology.

Once nucleated, dislocations can either exit the sample through the traction-free sides, annihilate with a dislocation of opposite sign when their mutual distance is less than $6b$ or become pinned at an obstacle. Point obstacles are included to account for the effect of blocked slip caused by precipitates and forest dislocations on out-of-plane slip systems that are not unassistedly described. They are randomly distributed over the slip planes with a constant density that corresponds on average, one source, either surface or bulk, that is accompanied by 8 randomly-distributed obstacles. It is worth mentioning that in this way the densities of sources and obstacles remains the same as the sample dimensions change, but there is a statistical preference towards always accompanying sources with obstacles in order to avoid statistical outlier behaviors. A dislocation stays pinned until its $PK$ force exceeds the obstacle-dependent value $\tau_{\rm obs}b$. The strength of the obstacles $\tau_{\rm obs}$ is taken to be $300$ MPa with 20\% standard deviation.

The simulation is carried out in an incremental manner, using a time step that is a factor 20 smaller than the nucleation time $t_{\rm nuc}=10\:$ns. At the beginning of every time increment, nucleation, annihilation, pinning at and release from obstacle sites are evaluated. After updating the dislocation structure, the new stress field in the sample is determined, using the finite element method to solve for the image fields \citep{vandergiessen1995}.

Our simulations are carried out for material parameters that are reminiscent of aluminum: $E = 70$ GPa, $\nu = 0.33$. We consider $50$ random realizations for each parameter case. 

After plastic yielding, abrupt plastic events can be directly defined as the sequence of consecutive stress drops --along the loading direction-- $\delta\sigma$, leading to the definition of equivalent energy release events 
\begin{equation}
S = \frac{\sigma_{\rm f}}{4E} \sum_{i\; \in\; {\rm event steps} }\delta\sigma_i. 
\label{eq:S}
\end{equation}
Here $\sigma_{f}$ donates the flow stress after plastic yielding, computed by averaging the stress between 2\% and 5\% total strain. Defined this way, $S$ is connected for each sample to the energy release during a single avalanche across the plastic flow regime.
Possible avalanche behavior is detected through power law tails of the event probability distributions $P(S)\sim S^{-\tau}{\cal P}(S/S_0)$ with $S_0$ being the cutoff of the distribution~\citep{salerno2013, Zaiser2007}.

\section{Strengthening size effects and statistical behavior of plasticity} 
\label{sec:basic}
\begin{figure}[t]
\centering
\subfigure[]{
\includegraphics[width=0.45\textwidth]{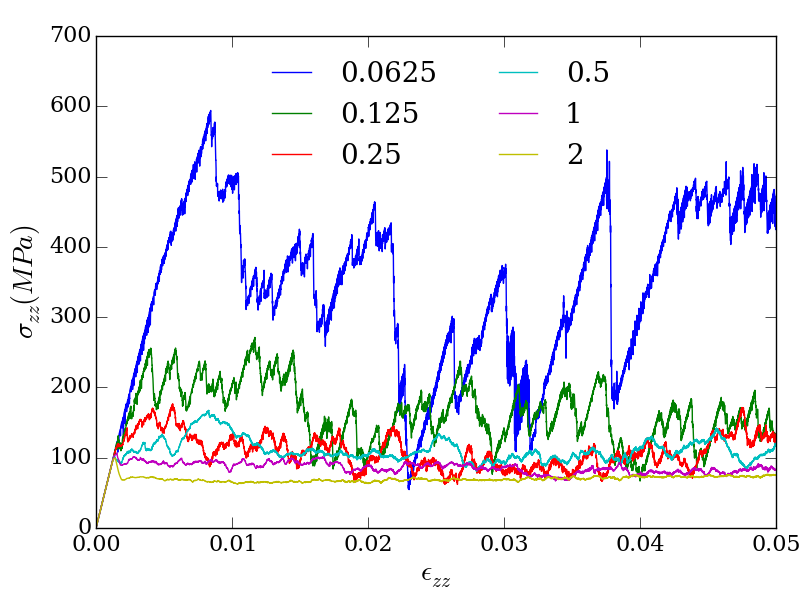}
}
\hspace{0.5cm}
\subfigure[]{
\includegraphics[width=0.2\textwidth, scale=0.5]{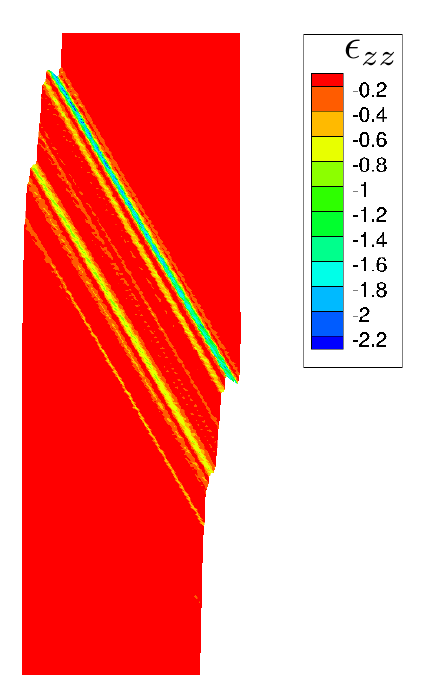}
}
\caption{(a) Axial stress--strain curves, $\sigma_{zz}$ vs $\epsilon_{zz}$. Strengthening and large stress drops emerge as $w$ decreases, with the width shown in the legend, in $\micron$, (b) uniaxial compression leads to characteristic slip patterns that are commonly observed in nanopillar uniaxial compression.}
 \label{fig:patterns}
\end{figure}

To investigate sample size effects, simulations were performed with different pillar diameters $w$ ranging from $0.0625\micron$ to $1\micron$ for a single slip system. Our simulations predict clear size effects in both the yield behavior and plastic flow regime. Fig.~\ref{fig:patterns}(a) shows typical examples of our predicted stress-strain curves, with characteristic strengthening and large flow stress fluctuations as $w$ decreases. Due to ideal strain-controlled loading, collective events emerge as stress-drops. A typical slip pattern (for $w=1\micron,\;\alpha=4$) that emerges, due to dislocation glide and escape from the free surface, in our simulations is shown in Fig.~\ref{fig:patterns}(b). The onset of plastic slip gives rise to plasticity-induced bending of pillars: this is discussed in detail in Appendix A. 

The bulk dislocation source density $\rho^{\rm{bulk}}_{\rm{nuc}}=60/\rm{\mu m}^2$ is chosen to be small enough such that the smallest sample width studied ({$w=0.0625\micron$}) contains no bulk dislocation sources.  This selection is consistent with the phenomenology of uniaxial nanopillar compression experiments, where it has been observed that dislocation starvation (no bulk dislocation sources) is present in samples with $w\lessapprox 100\nm$~\citep{Greer2011}. In this limit, in our model, single dislocations are generated by surface sources and placed at distance $10b$ from the free surface; this is approximately equivalent to considering a dislocation nucleation strength at the boundary to be $312$ MPa which is higher than the average obstacle strength. Moreover, this surface dislocation nucleation strength is approximately an order of magnitude larger than the average dislocation source strength in the bulk at larger widths.

\subsection{Yield stress and plastic flow stress fluctuations}

\begin{figure}[t] 
\centering
\subfigure[]{
\includegraphics[width=0.45\textwidth]{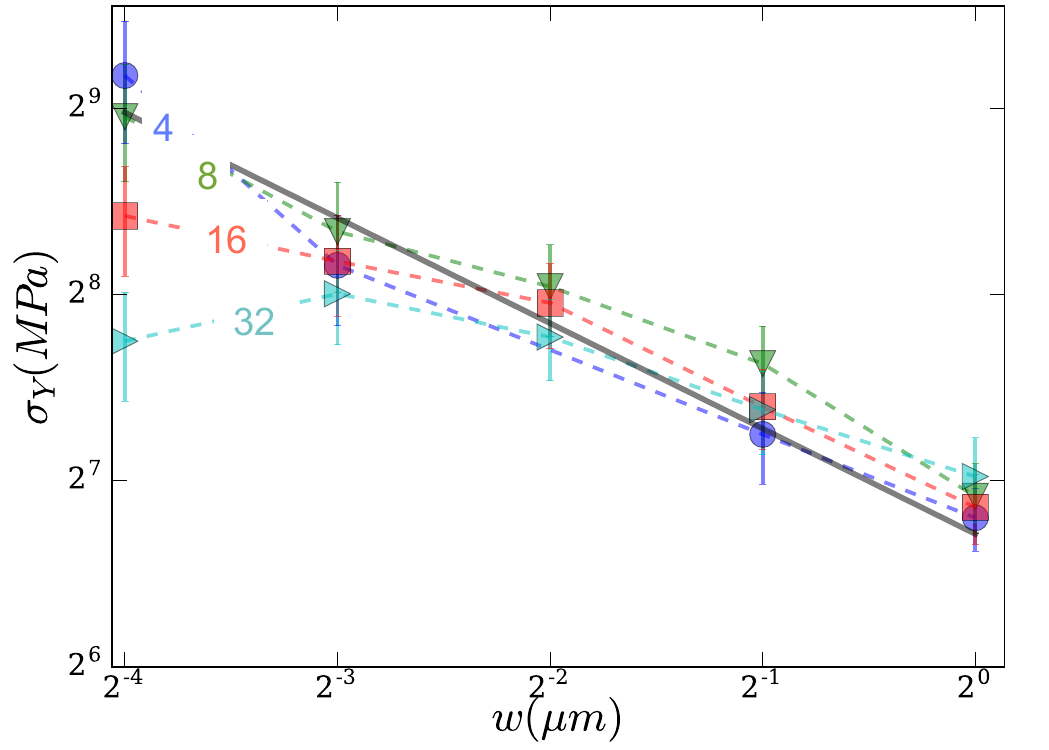}
}
\subfigure[]{
\includegraphics[width=0.45\textwidth]{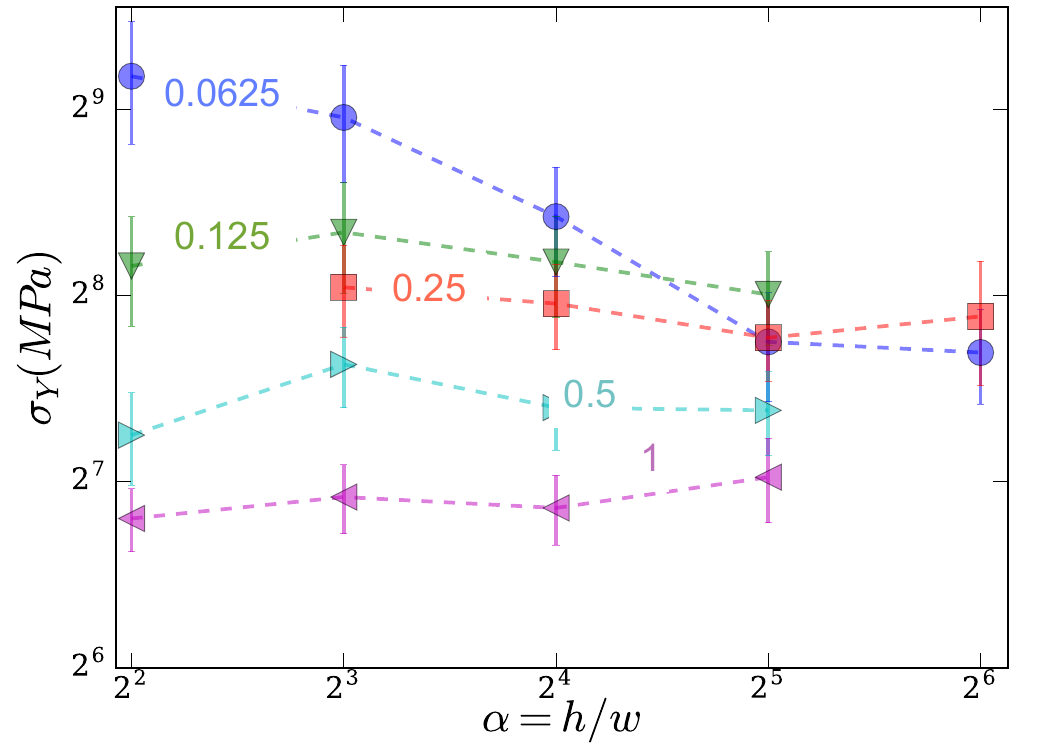}
}
\caption{(a) The width dependence of the yield stress. Different aspect ratios $\alpha$ are indicated by colored numbers. The fit to $\alpha=4$ is shown in bold black line, (b) the dependence of the yield stress on the aspect ratio $\alpha$.}
 \label{fig:size-effect}
\end{figure}

As shown in Fig.~\ref{fig:size-effect}, we identify distinct size effects in the yield strength, defined similar to engineering practice as the average stress between $0.1\%$ and $0.3\%$ plastic strain. 
Fig.~\ref{fig:size-effect}(a) shows that the yield strength $\sigma_Y$ decreases with increasing $w$. The fit to the aspect ratio $\alpha=4$ shows a clear power-law dependence $\sigma_Y \sim w^{-0.45}$ which is very close to experimental observations for small aspect ratios. In addition to a characteristic dependence on $w$, the sample strength depends on $\alpha$, as also identified in recent experiments~\citep{Senger2011, volkert2006,kiener2008}, decreasing strongly with a power law $\sigma_Y\sim \alpha^{-0.36}$ for small widths. In larger samples, this dependence is virtually absent, as shown in Fig.~\ref{fig:size-effect}(b). 

\begin{figure}[t!]
\centering
\includegraphics[width=0.5\textwidth]{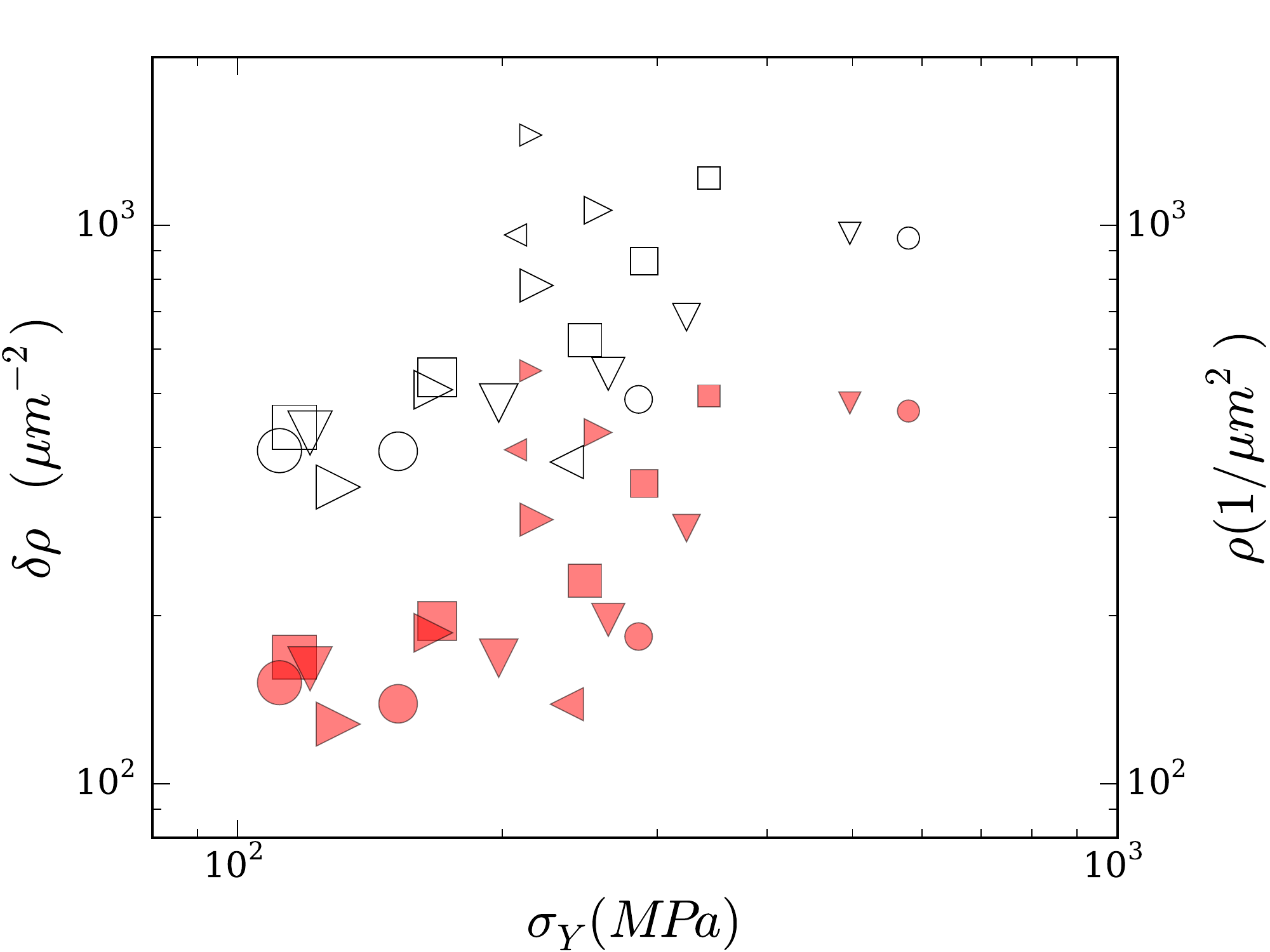}
\includegraphics[width=0.45\textwidth]{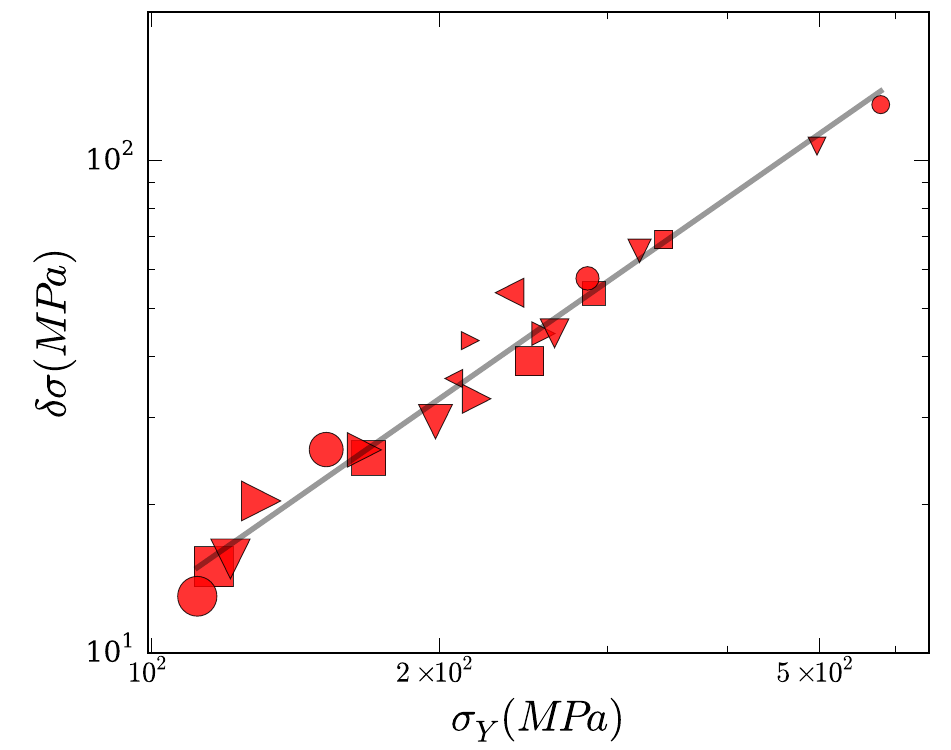}
\caption{
 {(a)} Scaling of dislocation density and dislocation fluctuations with yield stress.
The standard deviation (dark symbols) as well as the average (open symbols) of the dislocation density display an increasing trend with the yield stress. A number of aspect ratios are shown ($\filledmedsquare: \alpha=4$, $\filledmedtriangleleft: \alpha=8$, $\bullet: \alpha=16$, $\filledmedtriangledown: \alpha=32$, $\filledmedtriangleleft: 64$) with the symbol size proportional to the sample width: $w=0.0625, 0.125, 0.25, 0.5, 1.0\micron$. 
 (b) Flow stress fluctuations at $2\%$ strain as a function of $\sigma_{Y}$, for the dependence studied in Fig.~\ref{fig:size-effect}. Flow stress fluctuations at $2\%$ strain as a function of $\sigma_{Y}$, for the dependencies studied in left panel(
disks
$\bullet: \alpha=4$, 
down-pointing triangles
$\filledmedtriangledown: \alpha=8$, 
squares
$\filledmedsquare: \alpha=16$, 
right-pointing triangles
$\filledmedtriangleright: \alpha=32$, with marker size denoting increasing $w$). Strong correlation with $\sigma_{Y}$ according to $\delta \sigma \sim {\sigma_{Y}}^{1.84}$ (unbiased fit).
} 
 \label{fig:s2}
\end{figure}

The yield strength also increases with the average dislocation density (Fig.~\ref{fig:s2}(a) -- right vertical axis) and its fluctuations (Fig.~\ref{fig:s2}(a) -- left vertical axis).  The dislocation density fluctuations are defined through the standard deviation of the dislocation density ($\delta\rho=\int\sqrt{\rho(\epsilon)-<\rho>})d\epsilon$ at the region around the yield stress, a quantity that  demonstrates a strong variability even if the data points represent sample averages in $50$ random realizations. The dislocation density and its standard deviation increase with sample strength and decrease with sample width (being proportional to the markersize), but these correlations do not emerge as clearly from our simulations because of unresolved sample-to-sample fluctuations.

Distinct from the yield strength, the flow stress $\sigma_f$ is defined as the average stress between 2\% and 5\% total strain. As shown in Fig.~\ref{fig:s2}(b), flow stress fluctuations (defined as the standard deviation of $\sigma_f$) are strongly correlated to the yield strength with $\delta\sigma \sim {\sigma_{Y}}^{1.84}$. This concrete and novel prediction  forms the major signature of the {\it unassisted dislocation depinning} mechanism we propose. This mechanism may be tested in future experimental efforts and 3D-DDD simulations, and it clearly distinguishes nanopillar crystals from other materials which display abrupt plastic flow, such as bulk metallic glasses~\citep{Zhang2005}, but with no such correlations.

\subsection{Statistics of abrupt events}

\begin{figure}[t!]
\subfigure[]{
\includegraphics[width=0.45\textwidth]{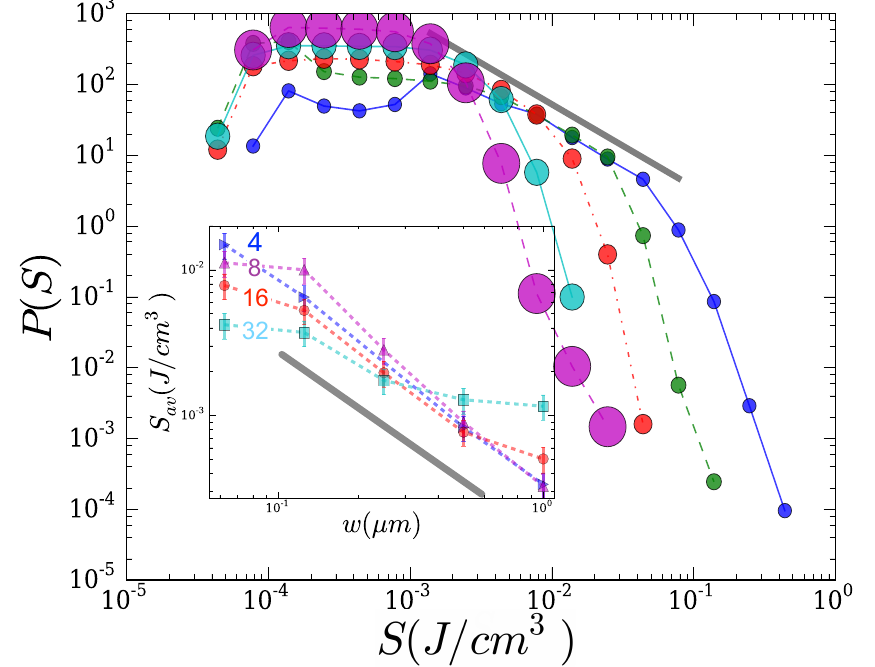}
}
\subfigure[]{
\includegraphics[width=0.48\textwidth]{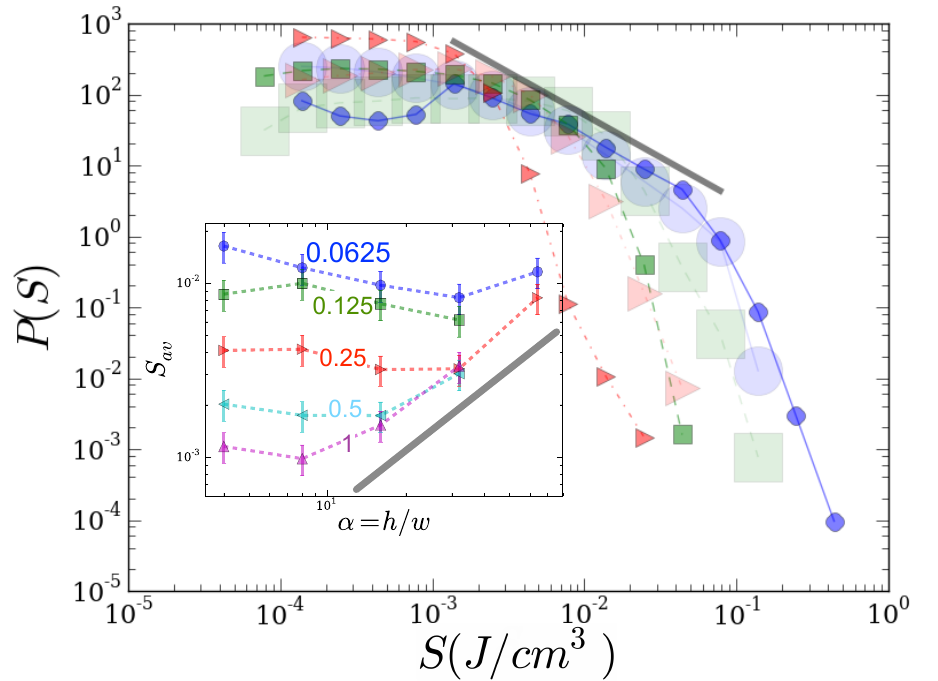}
}
\caption{
{Histograms of abrupt events and cutoff dependence}. (a) Width dependence of $P(S)$, demonstrating a clear power-law distribution as width decreases for $\alpha=4$ (here, symbol size reflects the width). In the inset, the average event size is shown as a function of $w$ for different aspect ratios. 
(b) Dependence of abrupt event statistics on pillar aspect ratio. Three different widths (
$\bullet: 0.0625\micron$, 
$\filledmedsquare: 0.25\micron$, 
$\filledmedtriangleright:1\micron$) are shown for two different aspect ratios $\alpha=4$ and $32$ (the symbol sizes follow the aspect ratio's magnitude for clarity).}
\label{fig:statistics}
\end{figure}

Possible avalanche behavior is detected through power law tails of the event probability distributions $P(S)\sim S^{-\tau}{\cal P}(S/S_0)$ with $S$ being defined in Eq.~(\ref{eq:S}) and $S_0$ signifying the large size cutoff of the probability distribution~\citep{salerno2013, Zaiser2007}. The onset of power-law behavior at decreasing width is clearly seen in Fig.~\ref{fig:statistics}a. Our findings are consistent with the existence of critical avalanche behavior ($S_0\rw\infty$) in the limit of infinitesimal width $w\rw0$: This observation is justified by the fact that as $w\rw0$, obstacles in the active slip planes are at infinitesimal distance and therefore, pile-up behavior is not possible. In this limit, dislocation dynamics becomes fully characterized by the depinning behavior from existing obstacles.
Further, the event distribution is characterized by an exponent $\tau=1.2\pm 0.2$ (the line $P(S)\sim S^{-1.2}$ is shown as a guide to the eye) while $S_0 \sim w^{-1}$.  The existence of power-law behavior in the asymptotically small width limit becomes apparent in samples with low aspect ratio, as shown in the inset of Fig.~\ref{fig:statistics}a (the line for the average event size $S_{av}\sim 1/w$ is shown as a guide to the eye). 
In Fig.~\ref{fig:statistics}(b), $P(S)$ is shown for three different widths ($0.0625$, $0.25$ and $1\micron$) and two aspect ratios ($4$ and $32$). $P(S)$ displays power-law behavior across aspect ratios (the line $P(S)\sim S^{-1.2}$ is shown as a guide to the eye), when the width is $0.0625\micron$; at larger widths, the distribution displays larger event sizes as the aspect ratio increases. This tendency is also seen in the behavior of $S_{av}$ (\cf inset of Fig.~\ref{fig:statistics}b), where independence of $\alpha$ is observed at small widths ($0.0625$ and $0.125\micron$), while there is an increasing trend with $\alpha$, $S_{av}\sim \alpha^{1}$, at large widths (with the line $S_{av}\sim \alpha$ as a guide to the eye). Similar behavior is observed also in the dependence of the distribution cutoff $S_0$ as function of width and aspect ratio. 
The complete behavior of the abrupt event energy release distributions as function of aspect ratio, width and obstacle strength points towards dynamical critical behavior~\citep{fisher1998} only in the limit of small widths and large aspect ratios.

\subsection{The role of surface dislocation sources for yield strength size effects and plastic burst distributions}
\begin{figure}[t!]
\centering
\subfigure[]{
\includegraphics[width=0.45\textwidth]{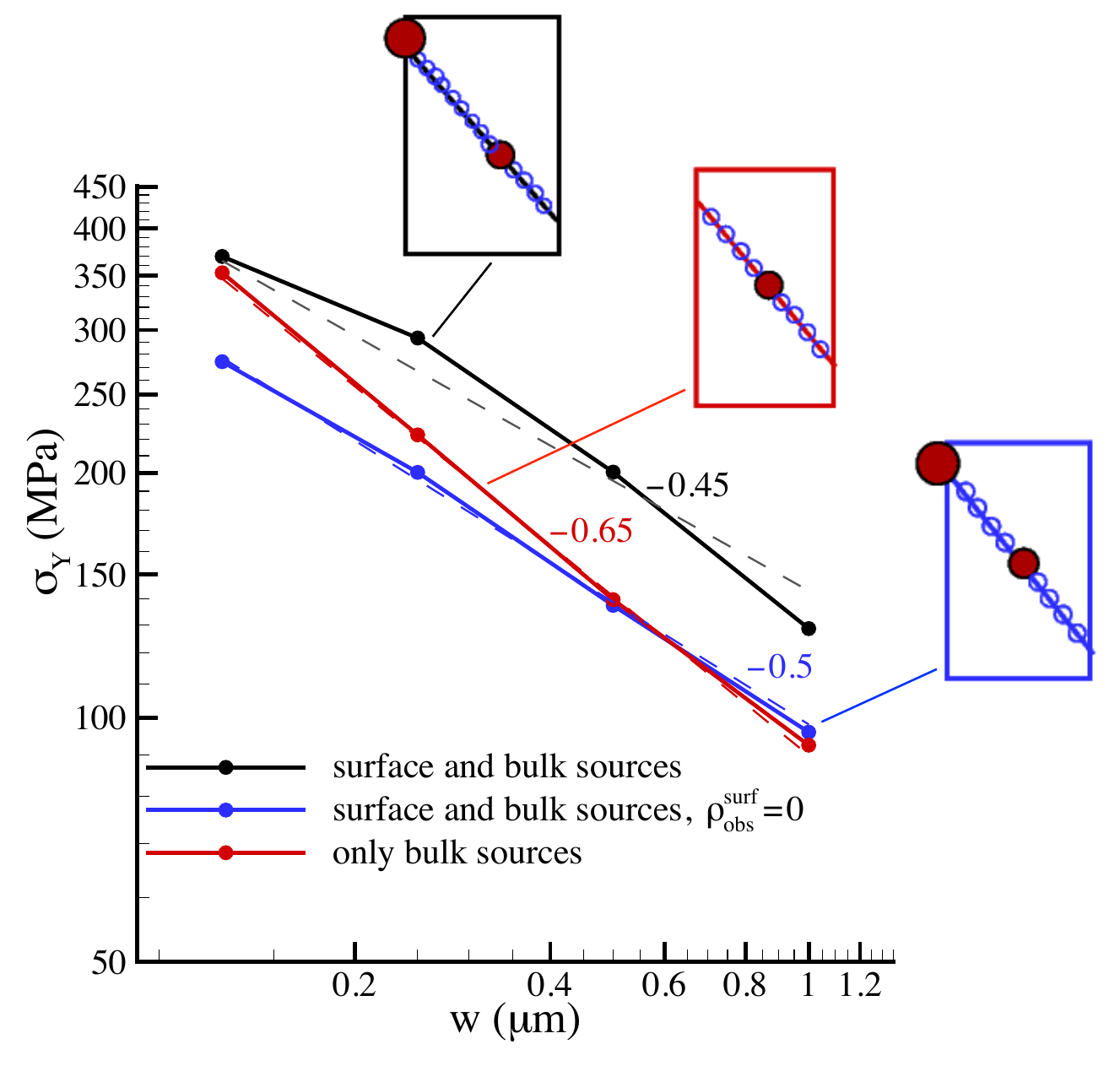}
}
\hspace{0.5cm}
\subfigure[]{
\includegraphics[width=0.45\textwidth]{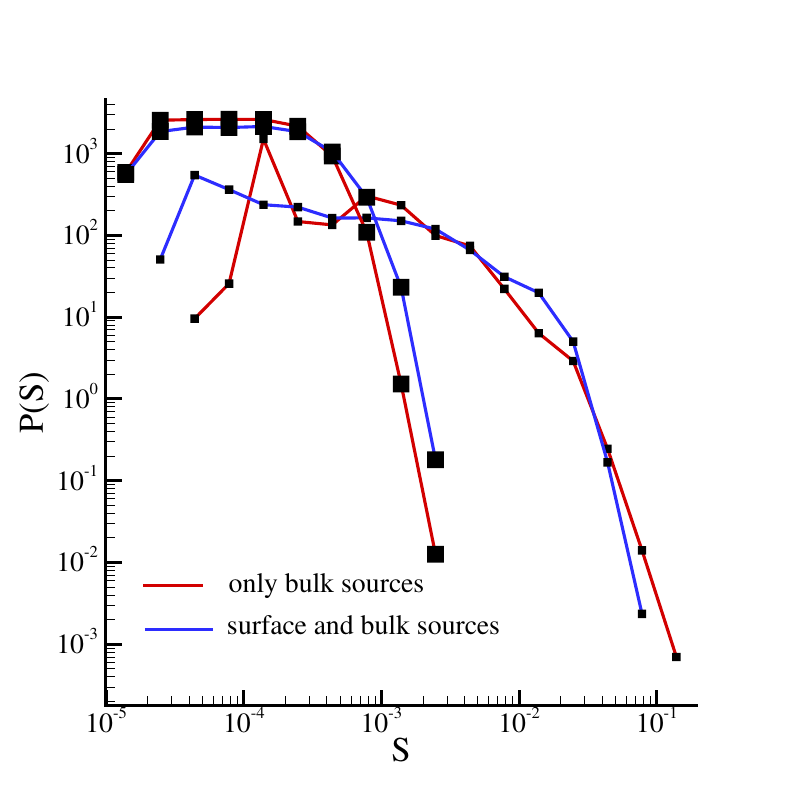}
}
\caption{Comparison of results between with and without surface sources and surface related obstacles. (a) Yield stress dependence on $w$, (b) statistics, larger symbol: $w=1\micron$, smaller symbol: $w=0.125\micron$.}
 \label{fig:compare-withoutsurface}
\end{figure}

In our model, surface sources for dislocations are designed to capture the primary experimental phenomenology of nanopillar uniaxial compression {by assuming the surface sources to have a much higher activation strength (312 MPa) than the }bulk sources (50 MPa). {Thus, the very existence of bulk sources naturally prevents surface sources from being activated.} 
Furthermore, while we imposed the surface source strength to be 312 MPa, molecular dynamics simulations  support the existence of even higher activation barriers (of the order of $\sim 1\rm{GPa}$),  which have been utilized in 3D-DDD simulations \citep{Ryu2015}. 
In practice, given the minimal character of our modeling, the phenomenology remains intact for any source activation strength larger than {the average obstacle strength of 300 MPa}. 

It is natural to ask whether strengthening size effects and  avalanche distributions are significantly affected by surface dislocation sources. Obviously, in our model, when $w=0.0625\micron$, the role of surface dislocation sources is the key to plastic deformation since there are no bulk dislocation sources. Thus, it is clear that for $w=0.0625\micron$ the role of surface dislocation sources is critical. However, is this true for $w>0.0625\micron$? 
 In order to answer this question, we perform simulations with all parameters intact as in the original model but with the surface dislocation sources removed. Then, our model contains only bulk sources and obstacles.  
In Fig.~\ref{fig:compare-withoutsurface}(a), if both surface and bulk dislocation sources and obstacles are present, the yield strength is overall {highest since the number of total obstacles is largest (given that each surface source is accompanied by eight obstacles)}. By retaining the surface sources but removing the surface-related obstacles ({blue} line), the yield strength $\sigma_{Y}$ {drop for all sizes, while} the power-law dependence is approximately unchanged ($-0.5$ compared to $-0.45$). {When subsequently also the surface dislocation sources are removed (red line), the yield strength $\sigma_{Y}$ remains the same on average for large $w$, because large samples have enough weak bulk sources that strong surface sources are not activated}. However, as $w$ decreases, {the exhaustion of bulk dislocation sources} leads to a stronger  strengthening size effect with power-law exponent $-0.65$. 

{Just like for strengthening, the absence of surface sources and/or obstacles has} an effect on avalanche distributions, but the difference in statistics has not been adequately resolved in our simulations, as is shown in Fig.~\ref{fig:compare-withoutsurface}(b). The statistics of the abrupt events for the small sample $w=0.125\micron$ appears to point to similar behavior with or without surface dislocation sources. Furthermore, the effect of surface sources on statistics appears to be approximately absent with increasing sample size, as shown for $w=1\micron$. 

Until the current point, all our simulations are characterized by weak dislocation sources and strong dislocation obstacles, and always the unstrained configuration is dislocation-free. Thus, it is natural to expect that when the sample size is large ($w\rw\infty$) the basic mechanism of yielding is pile-up dominated {(\cf \cite{curtin}). In this scenario, which we label as {\it assisted dislocation depinning}}, a weak dislocation source nucleates enough dislocations that can pile-up against an obstacle in order to form a large enough stress concentration that should move the leading dislocation across the obstacle. Naturally, this mechanism is fully functional at large $w$, where the {shear} yield stress is close to the average dislocation nucleation resolved shear stress ($50\rm{MPa}$) but much smaller than the average obstacle strength ($300\rm{MPa}$). In the regime of assisted dislocation depinning, plastic flow is smooth and yielding is size-independent. 

However, as the sample size decreases, assisted depinning is not anymore possible, since the space between nearby obstacles and sources decreases and therefore the available space in front of the obstacle is too small to generate a large enough pile-up stress. In this regime, dislocation pile-ups are ``incomplete" and {the yield strength} increases to levels required for {\it unassisted dislocation depinning} of individual dislocations through obstacles. This mechanism resembles analogous elastic depinning phenomena through high-disorder landscapes~\citep{fisher1998}. {Unassisted} depinning naturally leads to strengthening, but also it leads to strongly abrupt and stochastic plastic flow, given that the stresses required for dislocation dynamics to be activated are highly correlated and stochastically defined. Thus, it is intuitively clear that unassisted dislocation depinning {involves} a deep, intrinsic connection between the yield strength size effect and the stochastic character of post-yield plastic flow. 

Our main conclusion on the existence of this assisted/unassisted dislocation depinning is clearly associated with our choice of random distribution for dislocation sources and obstacles: Namely, as sample size becomes smaller and therefore the number of sources in the bulk decreases, all bulk sources are on different active slip planes (since we consider closely spaced ($10b$) available slip planes). Then, we can keep the bulk obstacle density fixed through distributing the required number of obstacles on {\it only} the active slip planes. This is an appropriate statistical approach of sampling obstacles, for otherwise --if for example, obstacles are distributed randomly across all slip planes \citep{deshpande2006}-- the behavior would be dominated by statistically outlier behaviors where slip planes exist where source activation takes place without obstacles. {In such a case, there is no dislocation pinning.}
\begin{figure}[t!]
\centering

\subfigure[]{
\includegraphics[width=.45\textwidth]{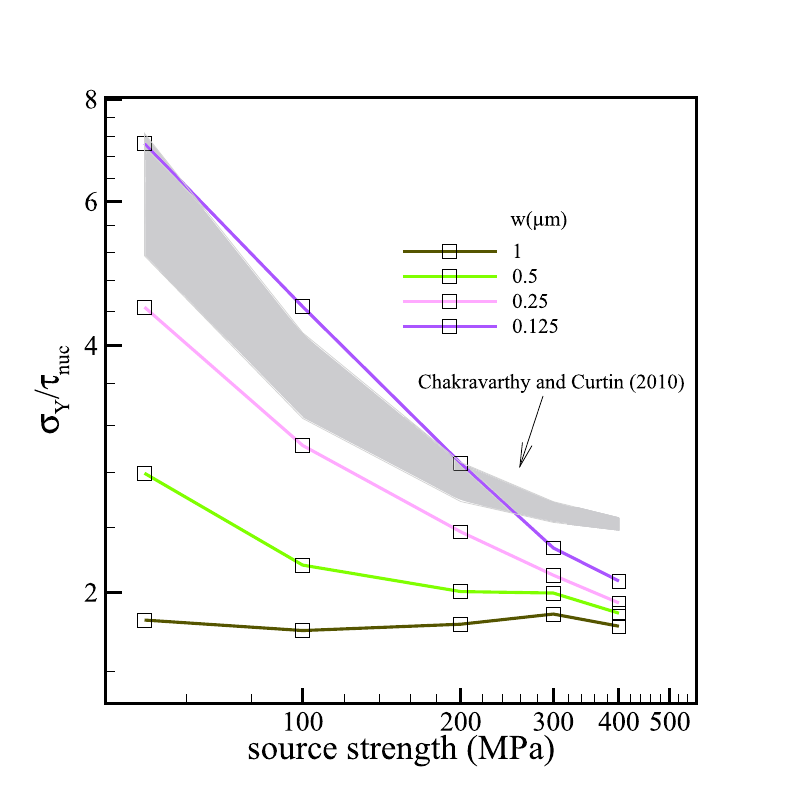}
}
\caption{Comparison of simulations results with { \cite{curtin}'s estimate (\ref{eq:curtin})} where the map is plotted based on the variations ($\pm$ SDV) of $\tau_{\rm obs}$  and $\tau_{\rm nuc}$ in our model.} 
 \label{fig:strengthening-source}
\end{figure}

The assisted/unassisted dislocation depinning crossover takes place {when} dislocation sources activate at much lower stress than the stress required to surpass obstacles. In this limit, there is a simple but highly approximate estimate of the yield strength of an edge dislocation pile-up configuration on a single slip plane~\citep{curtin}. {The estimate 
\begin{equation}
\label{eq:curtin}
\sigma_Y=M^{-1}\sqrt{\tau_{\rm nuc}^2 + 2A b \tau_{\rm obs}/L_{\rm obs}}
\end{equation} 
expresses a strong dependence on obstacle spacing $L_{\rm obs}$ and the  strength of obstacles, $\tau_{\rm obs}$ and of sources, $\tau_{\rm nuc}$ ($A$ is a fraction of the shear modulus, $M$ is the Schmid factor).}  
As {displayed in Fig.~\ref{fig:strengthening-source}, our model predictions and this estimate come close only for small sample width with weak} dislocation sources. The agreement in this limit ($w\rw0$) is justified by the fact that the model of \cite{curtin} does not take into account interactions with {dislocations on} neighboring slip planes which is highly possible for large samples. In our model,  when the sample width is small enough, active slip planes (at fixed dislocation source density) are at large mutual distance and therefore inter-slip interactions between dislocations become weak enough for single-slip-plane dislocation dynamics to {dominate}.

If the assisted/unassisted dislocation depinning crossover takes place in the weak-source (compared to obstacles) limit, it is natural to question what happens when source activation and obstacle strength are similarly valued. {This is what will be explored in the next session.}

\section{{Source} {\it vs.} obstacle strength and the role for yield strength, size effects and statistical behavior}
\label{sec:curtin}

\begin{figure}[b!]
\centering
\subfigure[]{
\includegraphics[width=.45\textwidth]{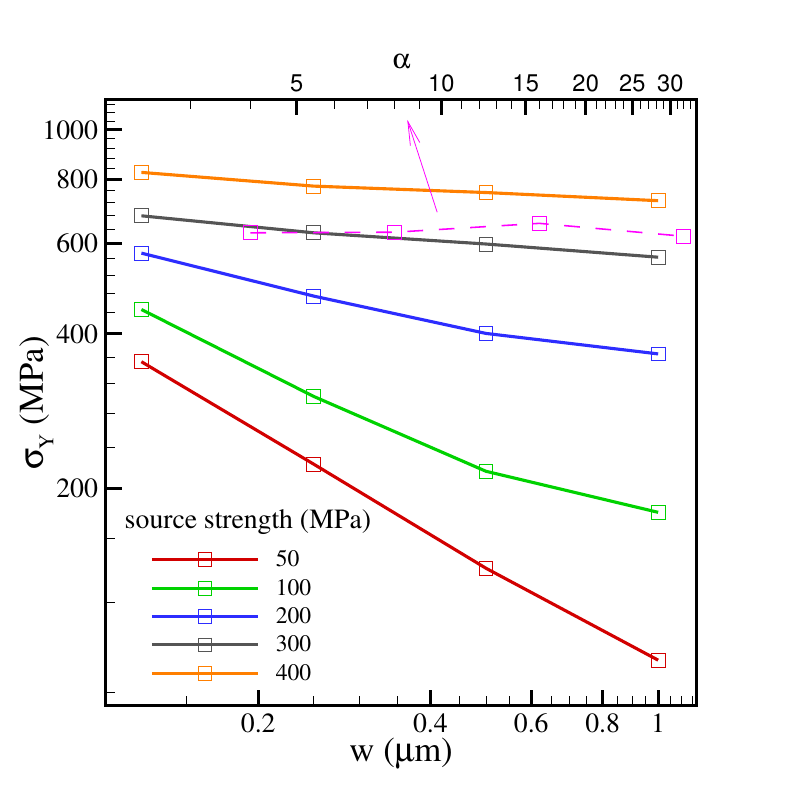}
}
\subfigure[]{
\includegraphics[width=0.45\textwidth]{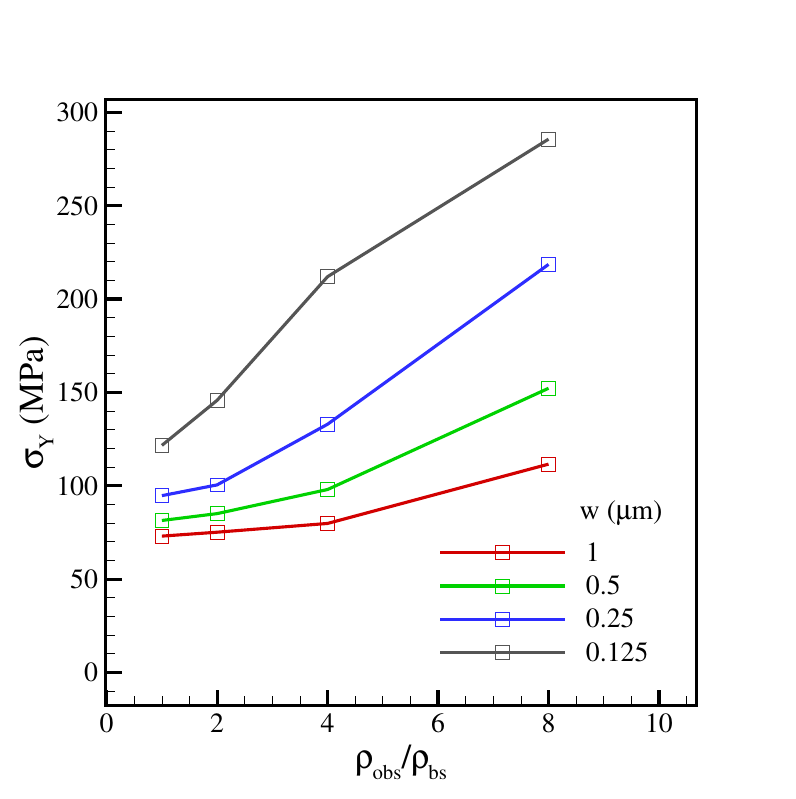}
}
\caption{ (a) Dependence of strengthening on the source strength. Different aspect ratio is shown for $w=0.25\micron$ when source strength is 300 MPa, (b) Dependence of $\sigma_{\rm{Y}}$ on the obstacle density for different sample width.}
\label{fig:obstacle-density}
\end{figure}

In this section, we present the results of extensive simulations to explore the dependence on the ratio of dislocation source activation barrier to obstacle strength. We keep the obstacle strength unchanged (300 MPa with $20\%$ SDV) and vary source strength from 50 MPa to 400 MPa. Our focus is on the dependence of strengthening and statistics on the variation of $\tau_{\rm nuc}/{\tau_{\rm obs}}$.  As it may be seen in Fig.~\ref{fig:obstacle-density}(a),  the average dislocation source strength increase leads to size-independent yield strength. This is a concrete signature that the yield strength size effect is strongly controlled by the assisted/unassisted dlslocation depinning mechanism we proposed. 
Obstacle strength as well as obstacle density have great influence on dislocation plasticity. 
In particular, the yield strength has a strong dependence on the obstacle density: the more obstacles results in more probable unassisted depinning. Furthermore, it appears that the obstacle density is the key factor (\cf Fig.~\ref{fig:obstacle-density}(b)) that determines the yield strength: for small widths, the increase of obstacle density leads to a large increase of the yield stress, while at large widths the increase of obstacle density does not influence the strength, as expected for a system that forms large dislocation pile-ups.

The role of obstacles is conditionally defined relative to the activity of dislocation sources. Our original model was defined in Section~\ref{sec:model} to compare well to experimental phenomenology of uniaxial nanopillar compression (see Section~\ref{sec:basic}). For it  assumes that bulk dislocation sources have a much lower activation stress than the strength of dislocation obstacles. In this way, the change of width $w$ leads to a crossover between assisted (at large $w$) --where pile-up behavior can take place-- and unassisted (at small $w$) dislocation depinning behavior, before dislocation starvation takes place at the ultra-small pillar regime.  However, it is clear that there is another fundamentally different regime of dislocation dynamics, is there strengthening or/and stochastic plastic behavior when the source strength becomes comparable to the obstacle strength? 
In the regime where bulk dislocation sources are as strong as the dislocation obstacles, yield strength size effects are not present anymore (\cf Fig.~\ref{fig:obstacle-density}(a)). But what is the corresponding statistical/stochastic behavior of the events?

\begin{figure}[t!]
\centering
\subfigure[]{
\includegraphics[width=0.45\textwidth]{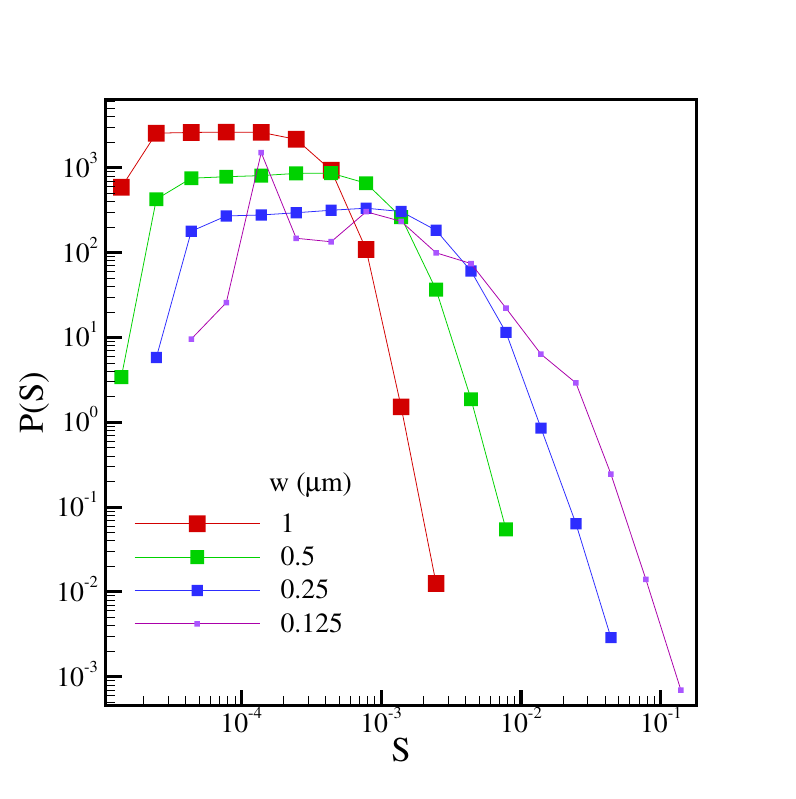}
}
\hspace{0.5cm}
\subfigure[]{
\includegraphics[width=0.45\textwidth]{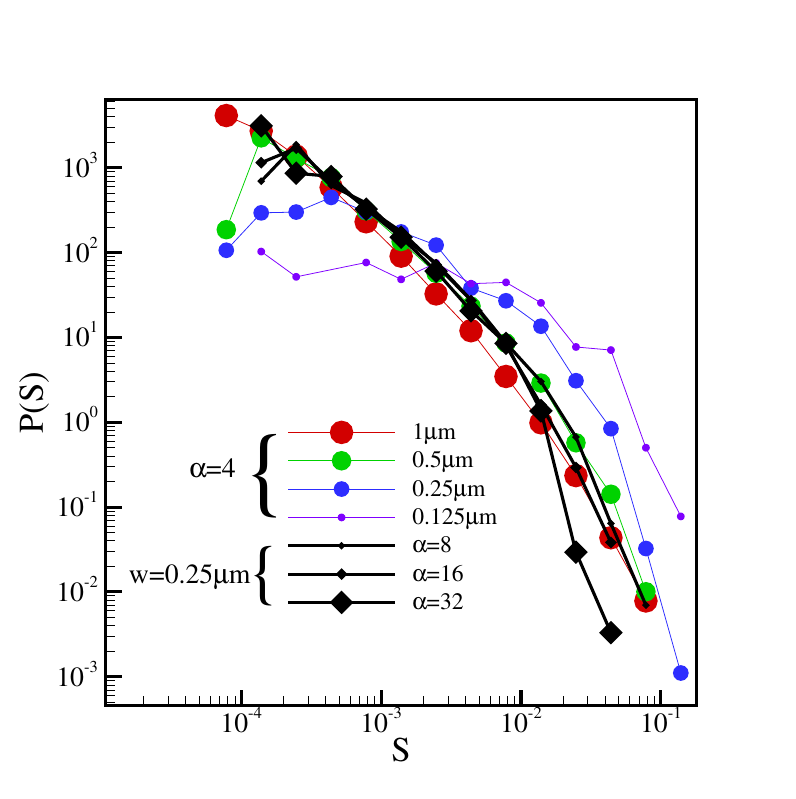}
}
\caption{ Events statistics for different source strength $\tau_{\rm{nuc}}$ (obstacle strength $\tau_{\rm{obs}}=300\rm{MPa}$ with 20$\%$ SDV). The size of the symbol stands for $w$ from 1$\micron$ (largest symbol) to 0.125$\micron$ (smallest symbol). (a) When $\tau_{\rm{nuc}}=50 \rm{MPa}$, (b) when $\tau_{\rm{nuc}}=300 \rm{MPa}$, blue lines are the statistics for $w=0.25\micron$ with different aspect ratios, symbol size stands for the value of aspect ratio.}
 \label{fig:statistics-differentsourcestrength}
\end{figure}

The statistics of plastic events is shown in Fig.~\ref{fig:statistics-differentsourcestrength} for the two cases of  dislocation sources with small/large activation strengths, keeping the obstacle strength fixed. When dislocation sources are weaker than the obstacles (\cf Fig.~\ref{fig:statistics-differentsourcestrength}(a)), the overall dependence on sample width $w$ is analogous to our main model and results discussed in Section~\ref{sec:model}, with a large sensitivity of the distribution cutoff on the sample width. When source strength becomes comparable to obstacle strength, statistics changes dramatically becoming approximately a pure power-law behavior, with a large-event cutoff seemingly independent of sample dimensions and geometry. 
Therefore, the statistical behavior of abrupt plastic events evidently appears universal across widths and aspect ratios (\cf Fig.~\ref{fig:statistics-differentsourcestrength}). This drastically new regime of source-driven deformation contains statistical noise that resembles acoustic emission measurements of crystalline materials~\citep{2, Weiss2003,weiss15}.
\begin{figure}[t!]
\centering
\subfigure[]{
\includegraphics[width=0.10\textwidth]{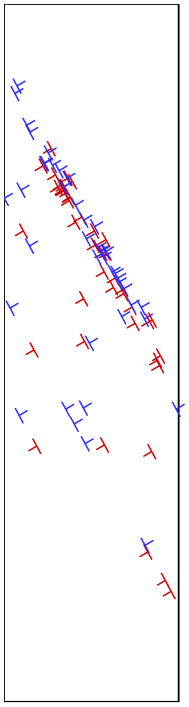}
}
\hspace{2cm}
\subfigure[]{
\includegraphics[width=0.2\textwidth,scale=0.8]{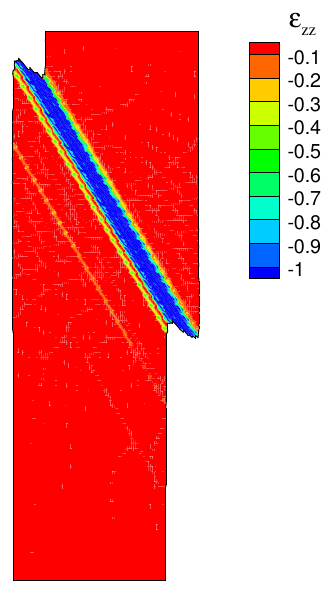}
}
\caption{The average source strength is 300MPa in a sample of $w=1\micron, \alpha=4$, (a) dislocation configuration at $5\%$ strain (plot in sample original configuration), (b) distribution of strain along the loading direction (plot in sample deformed configuration).}
 \label{fig:friction-strongsources}
\end{figure}

\begin{figure}[t!]
\centering
\subfigure[]{
\includegraphics[width=0.1\textwidth]{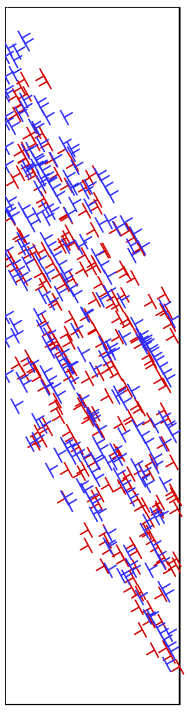}
}
\hspace{2cm}
\subfigure[]{
\includegraphics[width=0.2\textwidth]{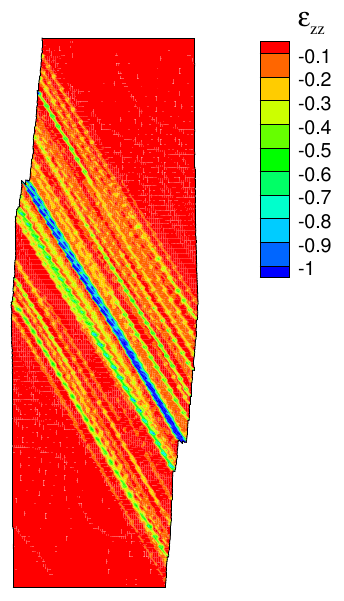}
}
\caption{The average source strength is 50MPa in a sample of $w=1\micron, \alpha=4$, (a) dislocation configuration at $5\%$ strain (plot in sample original configuration), (b) distribution of strain along the loading direction (plot in sample deformed configuration).}
 \label{fig:pileup-weaksources}
\end{figure}

The drastically different character of plastic events can also be traced in the spatial and temporal character of the dislocation motion during the energy release events. As shown in Fig.~\ref{fig:pileup-weaksources}, weaker sources lead to large, long-range inter-slip correlations (seen by the lateral strain correlations at large strains) -- facilitating avalanche events. Nevertheless, these events involve only a few dislocations at a time, consistent with experiments and other dislocation dynamics simulations~\citep{csikor2007, jaafar2015}. In contrast, when source strength is comparable to obstacle strength, events are traced in dominant dislocation pile-ups on a single slip plane. Long pile-ups accumulated during plastic flow generate long-range stress fields that lead to activation of nearby bulk dislocation sources. It is {noteworthy} that dislocation interactions lead to the fact that regions of strain localization are {associated with high dislocation density regions, as }shown in Fig.~\ref{fig:friction-strongsources}. The result is a self-organized dynamical critical behavior, analogous to self-organized earthquake faults~\citep{fisher1998}: the motion of few dislocations at the predominant slip plane drags along a large collection of nearby dislocations. This structural behavior is reminiscent of the coarse-slip-band (CSB) phenomenon~\citep{ar1977} which is believed to lie at the core of crystal plasticity instabilities~\citep{dr1996}.
This behavior should be contrasted to the homogeneous deformation in the case of weak sources (\cf Fig.~\ref{fig:pileup-weaksources}). Thus, we conclude that dislocation dynamics is dominated by the  yielding of a primary slip band with large dislocation density and large frustration between oppositely signed dislocations (\cf Fig.~\ref{fig:friction-strongsources}), a novel regime of dislocation friction that resembles coarse slip bands in bulk crystals~\citep{ar1977, dr1996}.

\section{Summary}

In this work, we presented a minimal model that aims to capture the basic aspects of uniaxial compression of nanopillars, including {yield strength size effects as well as} stochastic effects of post-yield plastic flow. 

When sources are much weaker than obstacles, strengthening with decreasing width is consistent with the experimentally observed scaling $\sigma_{Y}\sim w^{-0.45}$. Energy release events statistically become larger as width decreases, and the statistical distribution of events $P(S)$ acquires a power-law tail in the limit of small widths with an exponent $\tau=1.5\pm0.2$ in $P(S)\sim S^{-\tau}$. 
When dislocation sources are comparable in strength or stronger than the obstacles, the strength is virtually independent of width or aspect ratio while the statistical distribution of plastic events appears universal across width and aspect ratio scales with $\tau=1.9\pm0.2$. 

We demonstrated that small activation strengths (compared to dislocation obstacle strengths) for dislocation sources leads to {\it assisted dislocation depinning} at large widths that is strongly associated with dislocation pile-ups, while the mechanical behavior crossovers to an {\it unassisted dislocation depinning} mechanism at small widths, where each dislocation is required to jump over obstacles without being assisted by any dislocation pile-ups. The crossover that we identified naturally leads to strengthening that is consistent with the experimental phenomenology of BCC/FCC uniaxial nanopillar compression, but also it leads to the onset of critical avalanches in the limit of small widths, where obstacles on active slip planes are asymptotically close. Stochastic plastic flow fluctuations are strongly connected to the yield strength through an almost quadratic dependence $\delta\sigma_f\sim \sigma_{Y}^{1.84}$, a prediction that is ingrained to the nature of our unassisted dislocation depinning mechanism.

Furthermore, we demonstrated that large activation strengths for dislocation sources, and given that the unstrained samples are mechanically annealed, lead to the absence of size effects and {\it universal} plastic flow across sample widths and aspect ratios. We identified the reason of this behavior to originate in the onset of large dislocation frustration along a predominant slip band in the sample. Such crackling slip bands resemble the onset of coarse shear bands (CSBs) in crystal plasticity.

\section*{Acknowledgments}
We would like to thank  D. M. Dimiduk, C. F. Woodward, E. Lilleoden, S. Zapperi and P. Ispanovity for inspiring comments and discussions. This work has been supported through a VIDI Grant (NWO, SP) as well as a DOE-BES grant (SP) and a FOM grant (HS, EVdG). This work benefited greatly from the facilities and staff of the High Performance Computing Center at the Johns Hopkins University as well as the Zernike Institute for Advanced Materials at The University of Groningen, the Netherlands.


\section*{Appendix A}
\begin{figure}[ht!]
\subfigure[]{
\includegraphics[width=0.45\textwidth]{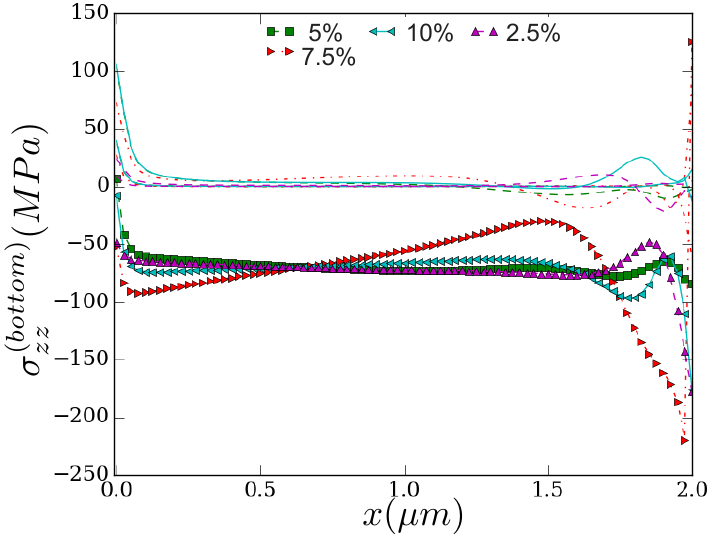}
}
\subfigure[]{
\includegraphics[width=0.45\textwidth]{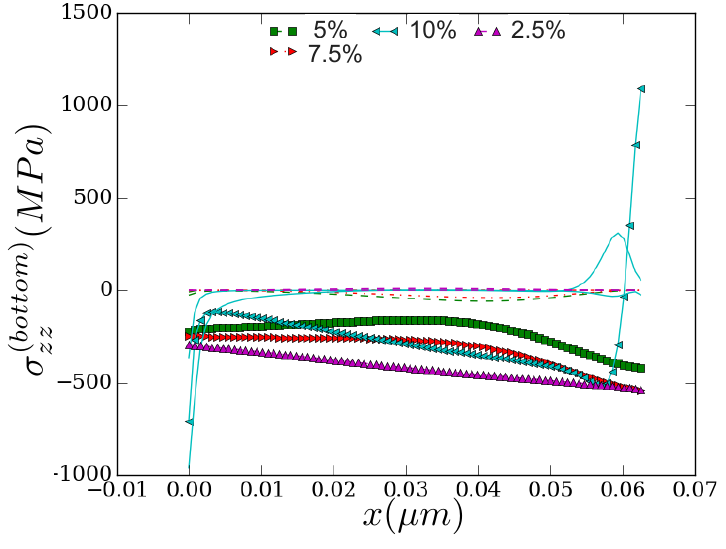}
}
\caption{
{\bf Bending effect in large (a, $w=2\micron$) and small samples (b, $w=0.0625\micron$) for fixed aspect ratio $\alpha=4$}, when the surface source density is high. The distribution of the stress components $\sigma_{zz}$ and $\sigma_{xz}$ along the bottom edge of the sample. The compressive stress $\sigma = -\sigma_{zz}$ (which is the experimentally relevant quantity,  measured through the indenter-load on pillars) display strong horizontal inhomogeneity as strain increases.
 }
 \label{fig:s9}
\end{figure}
Here, we investigate the bending effects that originate into the  elastic-plastic interplay during deformation. As the sample aspect ratio increases, it is natural to expect that the sample is more prone to bending, due to plastic slip, even though the prescribed loading is uniaxial compression. Buckling  evidently did not occur in any of our case studies, since: 
(1)  the total applied strain is  too small to cause such  an instability; (2) the onset of the instability should have become visible at large aspect ratios. We find strong stress inhomogeneity along the bottom surface of the sample, as demonstrated in Fig.~\ref{fig:s9}, given that our loading is purely strain-controlled. This stress inhomogeneity  develops also along the top surface of the sample, remains stochastic (its form varies across samples),  and is roughly independent of the width of the sample. Clearly the effect becomes dominant as the strain increases (from 1.25 to 5$\%$).

\section*{Appendix B}
In the main text, we present results with only one slip system, as shown in Fig.~\ref{fig:schematic}. Here, we verify strengthening (Fig.~\ref{fig:size-effect}(a)) as well events statistics (Fig.~\ref{fig:statistics}(a)) when 2 slip systems are used (the second slip system is added at an angle of $60$ degrees with respect to the first one (shown in Fig.~\ref{fig:schematic}), i.e., $150^\circ$ away from the loading direction ). 
\begin{figure}[ht!]
\centering
\subfigure[]{
\includegraphics[width=0.45\textwidth]{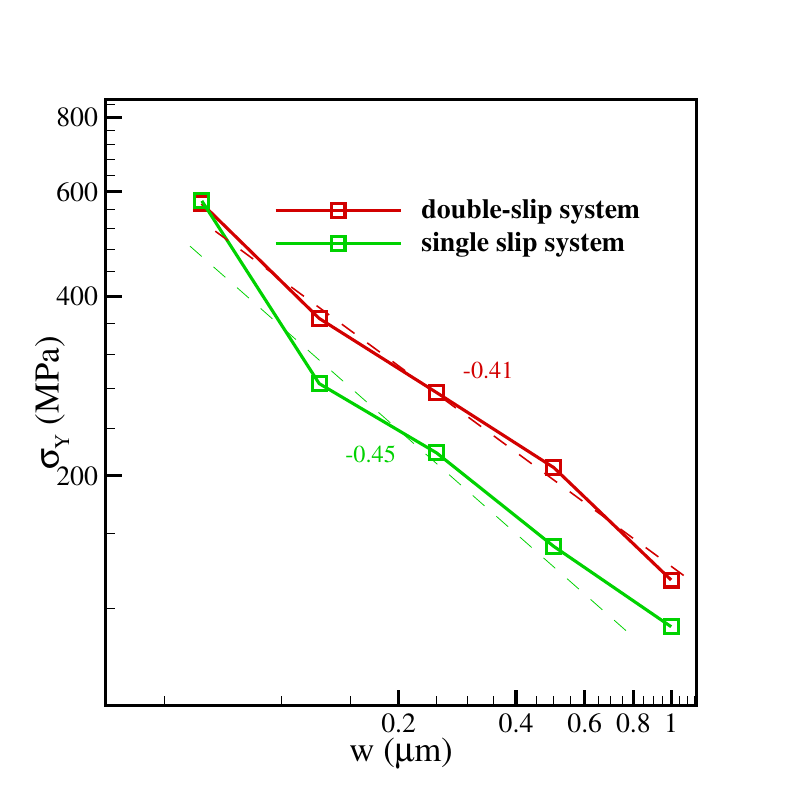}
}
\subfigure[]{
\includegraphics[width=0.42\textwidth]{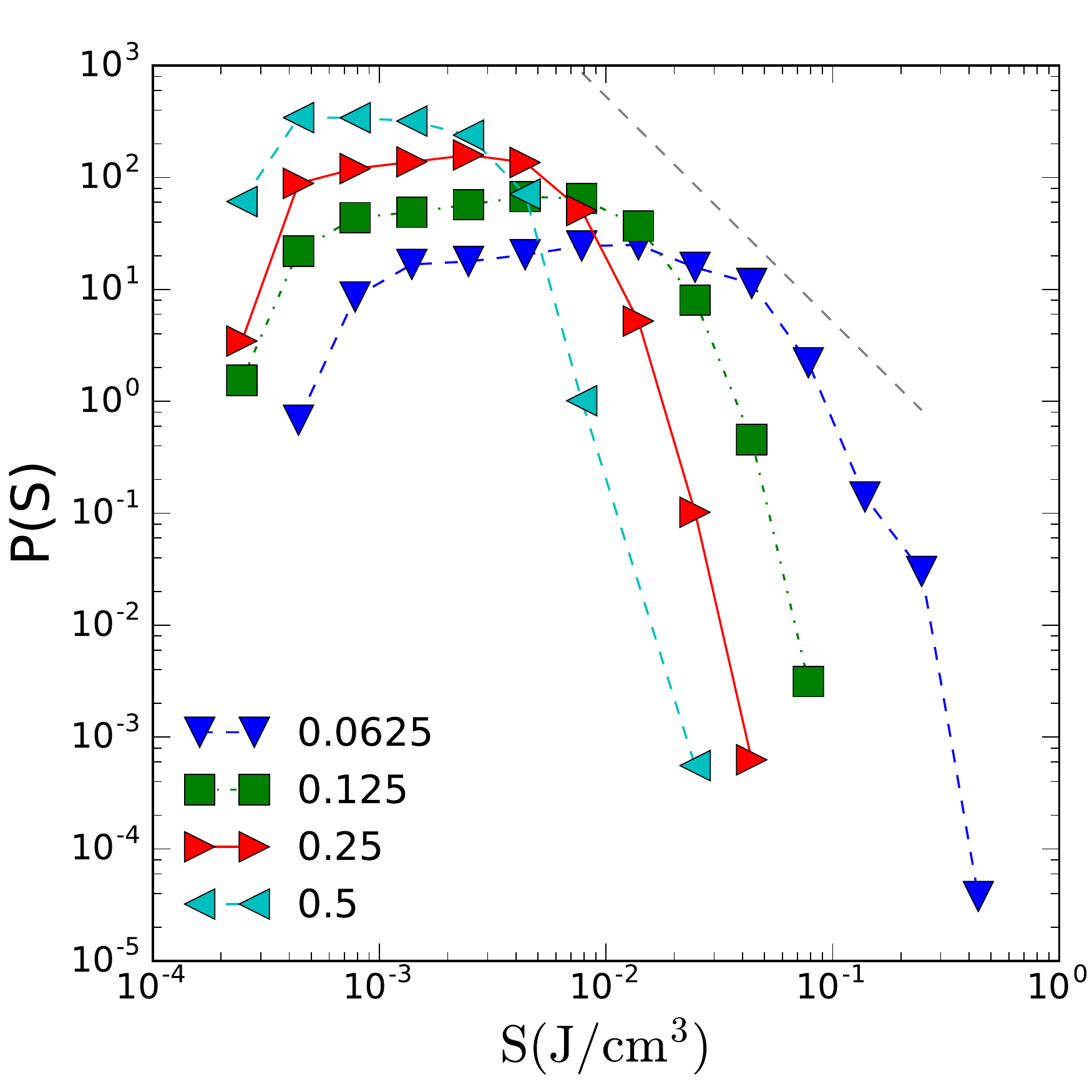}
}
\caption{(a) Dependence of strengthening on $w$. Dashed lines signify guides to the eye for power law strengthening with exponent labeled on the figure. (b) Statistics when $w$ changes from $1\micron$ (purple) to $0.0625\micron$ (blue) in the double-slip dislocation system, described in the text. The dashed line signifies an inverse-quadratic power-law and it just serves as a guide to the eye.} 
 \label{fig:2slip}
\end{figure}
As it can be seen in Fig.~\ref{fig:2slip}, there exist several main features that were discussed in the main text and are applicable in the case of a single slip system, such as the experimentally comparable yield strength size effect and the strong sensitivity of large event cutoff on sample width. Given that the number of surface dislocation sources is necessarily doubled, the inclusion of 50 samples appears to not completely resolve the strong statistical fluctuations in the distributions. However, given that the primary features of our work remain intact for the double-slip system, we focus solely on the single-slip system for computational convenience.  
\end{document}